\documentclass[a4paper,11pt]{article}
\pdfoutput=1 

\usepackage{jheppub} 

\usepackage[T1]{fontenc} 

\title{\boldmath Lorentz symmetry violation and the tunneling radiation of fermions with spin $1/2$ for Kerr Anti-de-Sitter black hole}

\author[a,1]{Zhi-E Liu,\note{Corresponding author.}}
\author[a]{Xia Tan,}
\author[a]{Yu-Zhen Liu,}
\author[a]{Bei Sha,}
\author[a]{Jie Zhang,}
\author[b]{and Shu-Zheng Yang}

\affiliation[a]{College of Physics and Electronic Engineering, Qilu Normal University, Jinan 250200, China}
\affiliation[b]{Department of Astronomy, China West Normal University, Nanchong 637002, China}

\emailAdd{zhieliu@163.com}
\emailAdd{adfyt@163.com}
\emailAdd{liuyz249@sina.com}
\emailAdd{shabei1234@163.com}
\emailAdd{zhangjie$\_$mail@126.com}
\emailAdd{szyangcwnu@126.com}

\abstract{We studied the correction of the quantum tunneling radiation of fermions with spin $1/2$ in Kerr Anti-de-Sitter black hole. First, the dynamic equation of spin $1/2$ fermions was corrected using Lorentz's violation theory. Second, the new expressions of the fermions quantum tunneling rate, the Hawking temperature of the black hole and the entropy of the black hole were obtained according to the corrected fermions dynamic equation. At last, some comments are made on the results of our work.}

\begin{document}
\maketitle
\flushbottom

\section{Introduction}	
Black holes are a special type of celestial bodies existing in the universe, including static black holes, stationary black hols and dynamic black holes. The curved space-time metric associated with a black hole possess is a solution to Einstein's field equation. Hawking found the black hole's thermal radiation, called Hawking radiation, by studying quantum effects near the black hole's event horizon ~\cite{Hawking1974,Hawking1975}. Hawking radiation effectively linked gravity theory, quantum theory and thermodynamic statistics physics, and inspired researchers to study the evolution of black hole thermodynamics ~\cite{Damour1976,Sannan1988,Zhao1991,Yang1995}. A theory that can really explain Hawking radiation is the quantum tunneling radiation theory, namely the event horizon of the black hole is taken as a barrier and due to the quantum tunneling effect, virtual particles inside the horizon have a certain probability to go through this barrier and be converted to real particles, then these real particles are radiated out of the black hole. Literature ~\cite{Kraus1995,Parikh2000,Hemming2001,Parikh2006,Akhmedov2006,Srinivasan1999,Shankaranarayanan2002,Medved2002,Iso2006,Zhang2006,Yang2005,Liu2019,Sha2020a,Sha2020b} used the quantum tunneling radiation method to study Hawking temperature and entropy of the black hole. The semi-classical theory proposed in literature ~\cite{Srinivasan1999,Shankaranarayanan2002} can deduce the Hamilton-Jacobi equation in curved space-time from the scalar field equation in curved space-time. Kerner and Mann et al. studied the tunneling radiation of Dirac particles using the semi-classical theory ~\cite{Kerner2008a,Kerner2008b,Criscienzo2008}. Lin and Yang proposed a new method to study the quantum tunneling radiation of fermions ~\cite{Lin2009b,Lin2009a,Lin2011,Yang2010}. Their method can also be used to study the quantum tunneling radiation of bosons. The results obtained in ~\cite{Lin2009b,Lin2009a,Lin2011,Yang2010} show that the Hamilton-Jacobi equation in curved space-time is the basic equation of particle dynamics, which reflects the inherent consistency between Lorentz symmetry theory and the Jamilton-Jacobi equation.

General relativity is a theory of gravity that cannot be renormalized, so several modified gravity theories have been proposed. Since researchers realized that Lorentz symmetry, the cornerstone of general relativity, may break at high energy cases, various gravity models based on Lorentz symmetry violation have been proposed ~\cite{Horava2009,Jacobson2001,Lin2014}. In principle, Lorentz symmetry violation theory can solve the problem of irrenormalization of gravity theory. In addition, some studies on Lorentz symmetry violation suggest that the dark matter theory may be just one of the effects of theoretical models of Lorentz symmetry violation ~\cite{Mukohyama2010}. In the fields of string theory, electrodynamics and non-abel theory, Lorentz symmetry violation has attracted extensive attention ~\cite{Colladay2007,Jackiw1999,Kostelecky1989}. In recent years, the Lorentz symmetry violational Dirac equation in flat space-time has been studied by introducing ether-like field terms, and the quantum correction of ether-like field terms was further studied ~\cite{Casana2011,Nascimento2015}. In this theory, the existence of ether-like field leads to the disappearance of Lorentz symmetry of the space-time. Therefore, properties that are inconsistent with Lorentz symmetry theory will emerge at high energy. These topics are worthy of study. On the other hand, for the dynamics of fermions in curved space-time, the modification of quantum tunneling radiation caused by ether-like field is also a subject worthy of further study when Lorentz symmetry violation is considered. At present, the quantum tunneling radiation of Dirac particles with ether-like field terms has been investigated merely in spherically symmetric black holes ~\cite{Pu2019}.

 In this paper, the quantum tunneling radiation of fermions is corrected in the axisymmetric stationary Kerr Anti-de-Sitter black hole by considering the ether-like field term. Our work is organized as follows: In section 2, according to Lorentz symmetry violation theory, the dynamic equation of fermions with spin 1/2 is derived for Kerr-Newman black hole. Section 3 will solve this dynamic equation and obtain the corrected physical quantities such as Hawking temperature and entropy of the black hole. The last section concludes our work.

\section{Lorentz symmetry violation theory and Dirac-Hamilton-Jacobi equation}

Reference ~\cite{Nascimento2015} studied the particle action and the Dirac equation in flat space-time based on Lorentz symmetry violation theory. Generalizing the ordinary derivatives in flat space-time to the covariant derivatives in curved space-time, and the commutation relation of Gamma matrices $\bar{\gamma}^{\mu}$ and $\bar{\gamma}^{\nu}$ in flat space-time to the commutation relation in curved space-time, we can get the Dirac equation with Lorentz symmetry violation for fermions with spin $1/2$ and mass $m$ in the curved space-time, namely ~\cite{Nascimento2015},

\begin{equation}
\{\gamma^{\mu}D_{\mu}[1+\frac{\hbar^2a}{m^2}(\gamma^{\mu}D_{\mu})^2]+\frac{b}{\hbar}\gamma^{5}+c\hbar (u^{\alpha}D_{\alpha})^2-\frac{m}{\hbar}\}\Psi =0,
\label{eq1}
\end{equation}
where $\gamma^{\mu}$ satisfies the following anti-commutation relations:

\begin{equation}
\gamma^{\mu}\gamma^{\nu}+\gamma^{\nu}\gamma^{\mu}=2g^{\mu\nu}I,
\label{eq2}
\end{equation}
\begin{equation}
\gamma^{5}\gamma^{\mu}+\gamma^{\mu}\gamma^{5}=0.
\label{eq3}
\end{equation}
Equation \eqref{eq2} turns to $\bar{\gamma}^{\mu}\bar{\gamma}^{\nu}+\bar{\gamma}^{\nu}\bar{\gamma}^{\mu}=2\delta^{\mu\nu}I$ and Eq.\eqref{eq3} turns to $\bar{\gamma}^{5}\bar{\gamma}^{\mu}+\bar{\gamma}^{\mu}\bar{\gamma}^{5}=0$ in flat space-time. In Eq.\eqref{eq1}, the general relativity derivative $D_{\mu}$ is defined as

\begin{equation}
D_{\mu} =\partial_{\mu}+\frac{i}{2}\Gamma^{\alpha\beta}_{\mu}\pi_{\alpha\beta}.
\label{eq4}
\end{equation}
In Eq.\eqref{eq1}, $a$, $b$ and $c$ are all small quantities. $u^{\alpha}$ is not a constant, but must meet the following condition:

\begin{equation}
u^{\alpha}u_{\alpha} = const.
\label{eq5}
\end{equation}
To solve the Dirac equation \eqref{eq1} for fermions with spin $1/2$, suppose
\begin{equation}
\Psi=\Psi_{AB}e^{\frac{is}{\hbar}}=\begin{pmatrix}
                                           A \\
                                           B \\
                                         \end{pmatrix}
e^{\frac{i}{\hbar}S},
\label{eq6}
\end{equation}
where $S$ is the Hamilton principal function. Substituting equations \eqref{eq4} and \eqref{eq6} into Eq.\eqref{eq1}, and noticing $\hbar$ is a small quantity, we can change Eq.\eqref{eq1} to

\begin{equation}
[i\gamma^{\mu}\partial_{\mu}S(1-\frac{a}{m^2}\gamma^{\alpha}\gamma^{\beta}\partial_{\alpha}S\partial_{\beta}S)
-cu^{\alpha}u^{\beta}(\partial_{\alpha}S\partial_{\beta}S)+b\gamma^5-m]\Psi=0.
\label{eq7}
\end{equation}
Using Eq.\eqref{eq2} we get

\begin{equation}
\gamma^{\alpha}\gamma^{\beta}\partial_{\alpha}S\partial_{\beta}S
=g^{\alpha\beta}\partial_{\alpha}S\partial_{\beta}S.
\label{eq8}
\end{equation}
According to Eq.\eqref{eq8}, Eq.\eqref{eq7} can be further converted to
\begin{equation}
\begin{split}
(i\gamma^{\mu}\partial_{\mu}S)\Psi
&=(1-\frac{a}{m^2}g^{\alpha\beta}\partial_{\alpha}S\partial_{\beta}S)^{-1}(cu^{\alpha}u^{\beta}\partial_{\alpha}S\partial_{\beta}S-b\gamma^5+m)\Psi \\
&=(1+\frac{a}{m^2}g^{\alpha\beta}\partial_{\alpha}S\partial_{\beta}S+\mathcal{O}(a^2))(cu^{\alpha}u^{\beta}\partial_{\alpha}S
\partial_{\beta}S-b\gamma^5+m)\Psi \\
&=[1+(\frac{c}{m}u^{\alpha}u^{\beta}+\frac{a}{m^2}g^{\alpha\beta})\partial_{\alpha}S\partial_{\beta}S-\frac{b}{m}\gamma^5]m\Psi\\
&=[1+(\frac{c}{m}u^{\alpha}u^{\beta}+\frac{a}{m^2}g^{\alpha\beta})\partial_{\alpha}S\partial_{\beta}S]m\Psi.
\end{split}
\label{eq9}
\end{equation}
The last step has used the condition $\frac{b}{m}\ll 1$. Multiplying both sides of this equation by $i\gamma^{\nu}\partial_{\nu}S$ , we get
\begin{equation}
(-\gamma^{\mu}\gamma^{\nu}\partial_{\mu}S\partial_{\nu}S)\Psi = [m^2+2(cmu^{\alpha}u^{\beta}+ag^{\alpha\beta})\partial_{\alpha}S\partial_{\beta}S]\Psi.
\label{eq10}
\end{equation}
With the help of Eq.\eqref{eq2}, the above equation becomes
\begin{equation}
[g^{\mu\nu}\partial_{\mu}S\partial_{\nu}S+2(cmu^{\mu}u^{\nu}+ag^{\mu\nu})\partial_{\mu}S\partial_{\nu}S+m^2]\Psi = 0.
\label{eq11}
\end{equation}
This is a matrix equation that has a nontrivial solution only if the determinant of its coefficient matrix associated with the fermions wave function $\Psi$ is zero, i.e.
\begin{equation}
g^{\mu\nu}\partial_{\mu}S\partial_{\nu}S+2(cmu^{\mu}u^{\nu}+ag^{\mu\nu})\partial_{\mu}S\partial_{\nu}S+m^2 = 0.
\label{eq12}
\end{equation}
Conducting some inferences and approximations can transform the above equation into
\begin{equation}
(g^{\mu\nu}+2cmu^{\mu}u^{\nu})\partial_{\mu}S\partial_{\nu}S+m^2(1-2a) = 0.
\label{eq13}
\end{equation}
From Eq.\eqref{eq1} to Eq.\eqref{eq13}, we not only get a new dynamic equation of Dirac particles, but also a deformed Hamilton-Jacobi equation (i.e. Eq.\eqref{eq13}) which is named as Dirac-Hamilton-Jacobi equation. $S$ is the Hamilton principal function whose expression depends on the selected coordinate system. In a stationary space-time, $S=S(t,r,\theta,\psi)$.

\section{Correction to tunneling radiation of spin $1/2$ fermions for Kerr Anti-de-Sitter black hole}
The Kerr Anti-de-Sitter black hole is a rotating black hole. The Vacuum solution of this black hole in Boyer-Lindquist coordinate system is expressed as ~\cite{Cardoso2004,Li2011}.

\begin{equation}
\begin{split}
ds^{2}=&-\frac{\Delta_{r}}{\rho^{2}}(dt-\frac{a}{\Xi}sin^{2}\theta d\phi)^{2}+\frac{\rho^{2}}{\Delta_{r}}dr^{2}+\frac{\rho^{2}}{\Delta_{\theta}}d\theta^{2}\\
&+\frac{\Delta_{\theta}sin^{2}\theta}{\rho^{2}}\left(adt-\frac{r^2+a^2}{\Xi}d\phi\right)^2,
\label{eq14}
\end{split}
\end{equation}
where
\begin{equation}
\begin{split}
&\rho^{2}=r^{2}+a^{2}cos^{2}\theta,\\
&\Delta_{r}=(r^{2}+a^{2})(\frac{r^2}{\ell^{2}}+1)-2Mr,\\
&\Delta_{\theta}=1-\frac{a^2}{\ell^{2}}cos^{2}\theta,\\
&\Xi=1-\frac{a^{2}}{\ell^{2}},
\label{eq15}
\end{split}
\end{equation}
 Here, $M$ is the mass of the black hole, $a$ is the angular momentum per unit mass ($a=\frac{J}{M}$) of the black hole, and $\ell$ is associated with the cosmographic constant by  $\ell^2=-\frac{3}{\Lambda}$. It can be seen from equations \eqref{eq14} and \eqref{eq15} that the non-zero entries of the covariance metric tensor are
\begin{equation}
\begin{split}
&g_{tt}=-\frac{\Delta_r}{\rho^{2}}+\frac{\Delta_{\theta}a^2sin^2\theta}{\rho^2},\\
&g_{\phi\phi}=-\frac{\Delta_{r}a^2sin^4\theta}{\rho^{2}\Xi^2}+\frac{\Delta_{\theta}(r^2+a^2)^2sin^2\theta}{\rho^2\Xi^2},\\
&g_{t\phi}=\frac{\Delta_{r}asin^2\theta}{\rho^2\Xi}-\frac{\Delta_{\theta}a(r^{2}+a^2)sin^2\theta}{\rho^2\Xi},\\
&g_{rr}=\frac{\rho^{2}}{\Delta_{r}},\\
&g_{\theta\theta}=\frac{\rho^{2}}{\Delta_{\theta}}.
\end{split}
\label{eq16}
\end{equation}
and the metric determinant and the non-zero entries of the inverse metric tensor are
\begin{equation}
\begin{split}
&g=-\frac{\rho^4sin^2\theta}{\Xi^2},\\
&g^{tt}=\frac{a^{2}sin^{2}\theta}{\rho^{2}\Delta_{\theta}}-\frac{(r^{2}+a^2)^2}{\rho^2\Delta_{r}},\\
&g^{\phi\phi}=\frac{\Xi^2}{\rho^{2}\Delta_{\theta}sin^2\theta}-\frac{\Xi^2a^2}{\rho^2\Delta_{r}},\\
&g^{t\phi}=\frac{\Xi a(r^{2}+a^2)}{\rho^2\Delta_{r}}-\frac{a\Xi}{\rho^2\Delta_{\theta}},\\
&g^{rr}=g^{11}=\frac{\Delta_{r}}{\rho^{2}},\\
&g^{\theta\theta}=g^{22}=\frac{\Delta_{\theta}}{\rho^{2}}.
\end{split}
\label{eq17}
\end{equation}
According to the null hyper-surface equation:

\begin{equation}
g^{\mu\nu}\frac{\partial F}{\partial x^{\mu}}\frac{\partial F}{\partial x^{\nu}}=0,
\label{eq18}
\end{equation}
we find that the event horizon of this black hole satisfies

\begin{equation}
\Delta_{r}=(r^2+a^2)(\frac{r^2}{\ell^2}-1)-2Mr=0.
\label{eq19}
\end{equation}
Equation \eqref{eq19} has two real solutions, denoted by $r_{H^-}$ and $r_{H^+}$, that represent inner and outer event horizon, respectively.\cite{Klemm1998,Dehghovni2002} 
Substitute Eq. \eqref{eq17} into Eq.\eqref{eq13}, and multiply the resulting equation by $\rho^2$, then we get the Hamilton principal function $S$ of spin $1/2$ fermions in Kerr Anti-de-Sitter black hole that satisfies

\begin{equation}
\begin{split}
&\Delta_r\left(\frac{\partial S}{\partial r}\right)^2-\frac{1}{\Delta_r}\left[(r^2+a^2)\frac{\partial S}{\partial t}+a\Xi\frac{\partial S}{\partial\phi}\right]^2
+\frac{1}{\Delta_{\theta}}\left[asin\theta\frac{\partial S}{\partial t}-\frac{\Xi}{sin\theta}\frac{\partial S}{\partial\phi}\right]^2\\
&+\Delta_{\theta}\left(\frac{\partial S}{\partial\theta}\right)^2
+2\rho^2cmu^{\mu}u^{\nu}\partial_{\mu}S\partial_{\nu}S+\rho^2m^2(1-2a)=0.
\end{split}
\label{eq20}
\end{equation}
To solve this equation, proper expressions should be given for $u^t$, $u^r$, $u^{\theta}$ and $u^{\phi}$. The principles of designing these expressions are: First, $u^{\alpha}$ is not a constant, but it should guarantee $u^{\alpha}u_{\alpha}=const$; second, $u^{\alpha}$ should be related to the properties of the metric tensor, since only the metric tensor can lift or lower the indices. Therefore, we use the following expressions for $u^t$, $u^r$, $u^{\theta}$ and $u^{\phi}$:
\begin{equation}
\begin{split}
&u^t=\frac{k_t}{\sqrt{g_{tt}}}=\frac{k_t\rho}{(\Delta_{\theta}a^2sin^2\theta-\Delta_r)^{1/2}},\\
&u^r=\frac{k_r}{\sqrt{g_{rr}}}=\frac{k_r(\Delta_r)^{1/2}}{\rho},\\
&u^{\theta}=\frac{k_{\theta}}{\sqrt{g_{\theta\theta}}}=\frac{k_{\theta}\rho}{(\Delta_{\theta})^{1/2}},\\
&u^{\phi}=\frac{k_{\phi}}{\sqrt{g_{\phi\phi}}}=\frac{k_{\phi}\rho\Xi}{sin\theta[\Delta_{\theta}(r^2+a^2)^2-\Delta_r a^2sin^2\theta]^{1/2}},
\end{split}
\label{eq21}
\end{equation}
where all of $k_t,k_r,k_{\theta},k_{\phi}$ are constants. $u^{\alpha}$ has the following property:
\begin{equation}
u^{\alpha}u_{\alpha}=k_t^2+k_r^2+k_{\theta}^2+k_{\phi}^2=const.
\label{eq22}
\end{equation}
Bringing Eq.\eqref{eq21} into Eq.\eqref{eq20}, and considering the fact that Eq.\eqref{eq19} holds at the event horizon of the black hole, we can get the dynamic equation of spin $1/2$ fermions at the event horizon of the black hole, that is,
\begin{equation}
\Delta^2_r\left|_{r\rightarrow r_H}\right. (1+2cmk^2_r)\left(\frac{\partial S}{\partial r}\right)^2\left|_{r\rightarrow r_H}\right.
-\left[(r_H^2+a^2)\frac{\partial S}{\partial t}+a\Xi\frac{\partial S}{\partial \phi}\right]^2 = 0.
\label{eq23}
\end{equation}
Then we get
\begin{equation}
\frac{\partial S}{\partial r}\left|_{r \rightarrow r_H}\right.=\pm\frac{r_H^2+a^2}{\Delta_r\left|_{r \rightarrow r_H}\right. \sqrt{1+2cmk^2_r}}
\left[\frac{\partial S}{\partial t}+\frac{a\Xi}{r_H^2+a^2}\frac{\partial S}{\partial \phi}\right].
\label{eq24}
\end{equation}
We can do variables separation in Eq.\eqref{eq29} by expressing the Hamilton principal function $S$ as
\begin{equation}
S=-\omega t+R(r)+\Theta(\theta)+j\phi,
\label{eq25}
\end{equation}
where $\omega$ is the energy of fermions with mass $m$, and $j$ represents the $\phi$ component of the generalized momentum of the radiated fermions. Since the curved space-time described by Eq.\eqref{eq14} is stationary and axisymmetric, $j$ is a constant. Note that $\frac{\partial S}{\partial t}=-\omega$ and $\frac{\partial S}{\partial \phi}=j$ can be easily deduced from Eq.\eqref{eq25}. Substituting Eq.\eqref{eq25} into Eq.\eqref{eq24} we get
\begin{equation}
\frac{\partial S}{\partial r}\left|_{r \rightarrow r_H}\right.=\frac{dR}{dr}\left|_{r \rightarrow r_H}\right.
=\pm\frac{r_H^2+a^2}{\Delta_r\left|_{r \rightarrow r_H}\right. \sqrt{1+2cmk^2_r}}
(\omega-\omega_0).
\label{eq26}
\end{equation}
where
\begin{equation}
\omega_0=\frac{a\Xi j}{r_H^2+a^2}
\label{eq27}
\end{equation}
is the chemical potential. Integrating the above equation from the inner side to the outer side of $r_H$ with the residue theorem, we obtain
\begin{equation}
\begin{split}
R&=\pm\int dr \frac{r_H^2+a^2}{\Delta_r\left|_{r \rightarrow r_H}\right. \sqrt{1+2cmk^2_r}}(\omega-\omega_0)\\
&=\pm i \pi \frac{r_H^2+a^2}{\Delta'_r\left|_{r \rightarrow r_H}\right. \sqrt{1+2cmk^2_r}}(\omega-\omega_0)
\end{split}
\label{eq28}
\end{equation}
where
\begin{equation}
{\Delta}'_r|_{r \rightarrow r_{H}}=2r_{H}(\frac{r_{H}^2}{\ell^2}+1)+2\frac{r_{H}}{\ell^2}(r_{H}^2+a^2)-2M
\label{eq29}
\end{equation}
Note that the equality $2Mr_{H}=(r_{H}^2+a^2)(\frac{r_{H}^2}{\ell^2}+1)$ holds due to Eq.\eqref{eq19}. The above equation becomes
\begin{equation}
\begin{split}
{\Delta}'_r|_{r \rightarrow r_{H}}&=2r_{H}(\frac{r_{H}^2}{\ell^2}+1)+2\frac{r_{H}}{\ell^2}(r_{H}^2+a^2)+\frac{1}{r_{H}}(r_{H}^2+a^2)(\frac{r_{H}^2}{\ell^2}+1)\\
&=\frac{2r^2_{H}(\frac{r_{H}^2}{\ell^2}+1)+2\frac{r^2_{H}}{\ell^2}(r_{H}^2+a^2)+(r_{H}^2+a^2)(\frac{r_{H}^2}{\ell^2}+1)}{r_{H}}.
\end{split}
\label{eq30}
\end{equation}
In Eq.\eqref{eq30}, $+$ and $-$ correspond to outgoing mode and incoming mode, respectively. Therefore, according to the quantum tunneling radiation theory, we can obtain the accurate expression of the quantum tunneling rate $\Gamma$ of Kerr Anti-de-Sitter black hole corrected by Lorentz violation theory, namely
\begin{equation}
\Gamma=exp[-2Im(R_+-R_-)]=exp(-\frac{\omega-\omega_0}{T_{H}}),
\label{eq31}
\end{equation}
where $T_H$ is the Hawking temperature at the event horizon of the black hole, with expression of
\begin{equation}
\begin{split}
T_{H}=&\frac{{\Delta}'_r|_{r= r_{H}}\sqrt{1+2cmk^2_r}}{4\pi (r^2_H+a^2)}\\
  =&\frac{{\Delta}'_r|_{r= r_{H}}}{4\pi (r^2_H+a^2)}(1+cmk^2_r+\cdots)\\
  =&T_h(1+cmk^2_r+\cdots).
\end{split}
\label{eq32}
\end{equation}
Here,
\begin{equation}
\begin{split}
T_h&=\frac{{\Delta}'_r|_{r= r_{H}}}{4\pi (r^2_H+a^2)}\\
&=\frac{2r^2_{H}(\frac{r_{H}^2}{\ell^2}+1)+2\frac{r^2_{H}}{\ell^2}(r_{H}^2+a^2)+(r_{H}^2+a^2)(\frac{r_{H}^2}{\ell^2}+1)}{4\pi r_{H} (r^2_H+a^2)}
\end{split}
\label{eq32p}
\end{equation}
 is Hawking temperature at the event horizon of Kerr Anti-de-Sitter black hole before our correction.

Obviously, the corrected Hawking temperature $T_H$ depends on $c$ and $k_r$, which represent the coupling strength and the radial component of the ether-like field respectively, and $T_H$ increases increasing values of them. Without the coupling parameter ($c=0$), we will recover the semi-classical Hawking temperature of Kerr Anti-de-Sitter black hole. Moreover, for the non-rotating ($a=0$) case, the temperature and its correction reduce to the case of Schwarzschild black hole.

\section{Entropy of Kerr Anti-de-Sitter black hole}
In the theory of black hole thermodynamics, an important physical quantity is black hole entropy. The correction of Hawking temperature will inevitably lead to the change of black hole entropy. According to black hole thermodynamics:
\begin{equation}
dM=TdS+VdJ+UdQ,
\label{eq33}
\end{equation}
where $S$ denotes the entropy, and the rotational potential $V$ and the electromagnetic potential $U$ are defined as
\begin{equation}
\begin{split}
U=&\frac{eQr_H}{r^2_H+a^2},\\
V=&\frac{aj}{r^2_H+a^2},
\end{split}
\label{eq34}
\end{equation}
where $e$ and $Q$ are the change of bosons and black hole, respectively, and $Q=0$ for non-charged Kerr Anti-de-Sitter black hole, resulting in $U=0$. At the event horizon $r_H$ the unmodified entropy satisfies
\begin{equation}
dS_h=\frac{dM-VdJ}{T_h}.
\label{eq35}
\end{equation}
After applying Lorentz violation, the entropy $S_H$ is corrected as
\begin{equation}
\begin{split}
S_H&=\int \frac{dM-VdJ}{T_H}\\
&=\int \frac{dM-VdJ}{T_h(1+cmk_r^2)}\\
&=\int(1-cmk_r^2+\cdots)dS_h\\
&= S_h-\int cmk_r^2dS_h+\cdots\\
\end{split}
\label{eq36}
\end{equation}
In the above calculation, we ignore small quantities of higher order. Since the quantum tunneling radiation of the black hole is a behavior of the radial direction of the black hole, only the $c$ term will affect the tunneling radiation of Dirac particles in the black hole. As can be seen from Eq.\eqref{eq32}, if the coupling parameter $c$ is positive, the Lorentz symmetry violation will, contrary to the case of Hawking temperature, decrease the entropy of Kerr Anti-de-Sitter black hole.

\section{Conclusion}

The main work of this paper is to correct the quantum tunneling radiation of spin $1/2$ fermions in the stationary kerr Anti-de-Sitter space-time with Lorentz violation theory. It is quite complicated to correct the dynamic equation of Dirac particles in curved space-time based on Lorentz symmetry violation theory. However, the results obtained in this paper are satisfactory. By studying equations \eqref{eq1} and \eqref{eq13}, we can simplify the complex calculation. The Hamilton principal function of Dirac particles is obtained from Eq.\eqref{eq13}, then the tunneling rate of the black hole, the Hawking temperature at the event horizon of the black hole and the entropy of the black hole are obtained after correcting. These research results are valuable for further research on the quantum gravity theory and the thermodynamic evolution of black holes.

In short, the modification to particle dynamics equations in curved space-time based on Lorentz violation theory and related problems are topics worthy of in-depth exploration. In recent years, it is believed that Lorentz symmetry will be broken at high energy scales. Therefore, the quantum field equation affected by Lorentz violation has been successively studied. These studies, including our research on particle dynamics equations in curved space-time, can provide theoretical support for research on quantum gravity. In curved space-time, the simple but special properties of the black hole surface may become a powerful tool for verifying Lorentz violation theory in the future. We will continue our research in this area.

On the other hand, with the advance of science and technology, in 2017, the gravitational waves generated by the merger of two black holes have been directly observed for the first time. This means that the gravitational radiation from the vicinity of the event horizon of the black hole will carry information of the extreme gravitational field, which will be of great interest to researchers. We will also carry out relevant research in the future.

\acknowledgments

This work is supported by the National Natural Science Foundation of China (No.11573022,11273020), and the Natural Science Foundation of Shandong Province, China (No. ZR2019MA059).


\end{document}